\documentclass[epj]{webofc}
\usepackage[varg]{txfonts}   
\wocname{EPJ Web of Conferences}
\woctitle{New Frontiers in Physics 2014}
\begin{document}
\selectlanguage{english}
\title{The Potential of Continuous, Local Atomic Clock Measurements for Earthquake Prediction
and Volcanology}
%
%

\author{Mihai Bondarescu \inst{1,2}\fnsep\thanks{\email{bondarescu.mihai@gmail.com}} \and
        Ruxandra Bondarescu, \inst{3} 
        Philippe Jetzer\inst{3} \and
        Andrew Lundgren\inst{4}
}

\institute{Physics Department, University of Mississippi, Mississippi, USA
\and
           Departamentul de Fizica, Universitatea de Vest, Timisoara, Romania
\and
          Department of Physics, University of Z\"{u}rich, Z\"{u}rich, Switzerland
           \and Albert Einstein Institute, Hannover, Germany
          }

\abstract{
Modern optical atomic clocks along with the optical fiber technology currently being developed can measure the geoid, which is the equipotential surface that extends the mean sea level on continents, to a precision that competes with existing technology. In this proceeding, 
we point out that atomic clocks have the potential to not only map the sea level surface on continents, but also look at variations
of the geoid as a function of time with unprecedented timing resolution. The local time series of the geoid has a plethora of applications. These
include potential improvement in the predictions of earthquakes and volcanoes, and closer monitoring of ground uplift in areas
where hydraulic fracturing is performed. Detailed results follow in \cite{us}.
}
\maketitle

\section{Introduction}
\label{intro}
On the global level, the geoid, which is an equipotential surface that closely reproduces
the global mean sea surface and extends to continents, is known to 30-50 cm with relatively
poor resolution as a function of space and time. Currently,
applied techniques that provide both high lateral resolution and accuracy are based on 
indirect approaches. Gravity measurements from orbiting satellites such as the
Gravity Recovery and Climate Experiment (GRACE) \cite{vonFrese, Tapley} provide information on the geoid
through integration of the observed gravity field. More recently, gradiometry data from
the Gravity Field and Steady State Ocean Circulation Explorer (GOCE) \cite{Paila, Pailb, Pavlis} became available.
To obtain the most accurate geoid models, twofold integration of gradiometry measurements
and calibration based on land measurements is performed.  Satellite measurements
suffer from poor spatial resolution that is of the order of the distance between the satellite
and Earth ($\sim 400$ km) and measure the attenuated gravitational field of the Earth at
the location of the satellite, which suffers from aliasing due to numerous superimposed
effects.

Atomic clocks are emerging as a new tool to measure the geoid locally on the Earth. The best 
clocks to date map changes of the geopotential of the Earth with an accuracy of
the order of $1$~cm over an integration period of a few hours \cite{Hinkley2013, Bloom2014}. In \citep{Bondarescu2012}
we argued that clocks in conjunction with gravimeters could be used to calibrate and add detail
to satellite maps, and could resolve underground structures with a radius of $\sim1.5$ km for a
$20\%$ density anomaly.  In this proceeding we argue that atomic clocks compete 
with precision of differential GPS, which can measure the uplift of the ground. Clocks provide an 
independent measurement that complements those by GPS receivers. The GPS clocks are much less 
accurate than the ground clocks and differential GPS measurements are limited by a number of effects 
such as the ionosphere delay and the distance to the base-station. They often require long integration times 
over periods of one year or more. Comparison between ground clock networks and the GPS clocks 
are currently underway.

In this proceeding we argue that atomic clocks could measure smooth alterations of the geopotential of the Earth on the sub-cm level.
 Additionally, measurements by local clocks in conjunction with measurements by quantum gravimeters in volcanic areas 
 would determine whether the ground uplift is correlated with mass redistribution, since the two scale differently with
 the distance from the source. A sequence of earthquakes that lead to a gradual change in elevation over a period of a few 
 days could be associated with magma movements underground, and potentially with future eruptions. 
 
 The slow down of time in the vicinity of heavy objects is a consequence of Einstein's theory of general relativity. 
 Massive objects curve space-time, and slow down time. An observer outside the horizon of a black hole, sees 
 time stopping all together at the black hole horizon. On a neutron star, clocks tick at about half their rate on Earth. 
 Similarly, clocks that are closer to Earth tick slightly slower than clocks that are further away.  Furthermore, sources
 below ground also affect the clock rate of local observers. If the magma chamber of a volcano fills up, the tick
 rate of the clock on the volcano will slow down. There may be a delay between the magma chamber filling up and 
 the ground uplift, which future clock networks could observe. Such an observation will likely require frequency stabilities
 beyond what is available today unless the clock is on a supervolcano where larger quantities of magma of volume
 $ \sim 1$ km$^3$ or greater could be shifting underground.
   
  Atomic clocks can also be used to measure the tidal response of the Earth on the local level. Very precise 
  measurements of tidal uplift may resolve changes in the deformability of the crust, a possible precursor to earthquakes.
  Unlike the GPS data, local clock measurements are not affected by atmosphere loading, passing clouds or 
  other atmospheric perturbations.

%
%

%
 


\section{Brief Overview of Atomic Clocks}
Progress in atomic clock technology already has had tremendous impact on our every day life.
The frequency stability of atomic clocks has been improving at a rate of about a factor of 10 every
decade for the past 60 years. After the discovery of the femtosecond laser frequency comb, which enabled
the counting of the oscillations of optical atomic transition, optical clocks have been
 improving at an even more rapid rate \cite{Poli2013}. Clocks are used to define both the meter and the second. 
Since 1983, the meter has been defined as the length of the path travelled by light in a vacuum
in $1/299 792 458$ of a second. This means that already in 1983 we were better at measuring 
1 part in $10^9$ of a second than at measuring the length of an object. 

A global fiber-link network capable of disseminating ultra-stable frequency signals is being built throughout Europe and
in the US. This network will connect clocks across the Earth, and provide continuous, accurate measurements of the tick rate 
of the clocks as a function of time. The clocks themselves are becoming more stable. The Physikalisch - Technische Bundesanstalt (PTB) group aims to operate an optical clock for one year without human intervention.
Atomic clocks measure changes in the gravitational potential, and provide the most direct way to determine the geoid of the Earth. 
The geoid is the equipotential surface of constant tick rate that extends the mean sea level. The best optical clocks to date could measure
the geoid to the cm-level over an integration period of about $7$ hours \cite{Hinkley2013, Bloom2014}. Eventually, ultra-performant optical clocks are expected to be hosted at observatories around the Earth, and connected via special channels in the fiber link used for the internet.

Optical clocks use optical transitions that have a much higher relative line-width ($Q = f/\Delta f \sim 10^{15}$ with $f$ being the frequency 
of the resonance) than microwave transitions ($Q \sim 10^{10}$) used for microwave clocks. The frequency stability of atomic clocks is 
typically quantified by the Allan variance, which for the best optical clocks to date is
$\sigma_{\rm today} \sim 3 \times 10^{-16}/\sqrt{\tau{\rm/sec}}$ \cite{Hinkley2013, Bloom2014}.  Today's best clocks are optical clocks whose frequency
stability is within a factor of $2-3$ of the Quantum Projection Noise (QPN) for their respective transitions, where $\sigma_{\rm QPN} \sim Q^{-1}N^{-1/2}$ 
with $N$ the number of trapped ions. This noise is a fundamental limit that can be lowered only by considering a different atomic transition.
Several avenues that would result in another dramatic improvement in atomic clocks are being pursued by experimental groups around the world. 
One possibility is to choose higher transition frequencies with low inherent environmental sensitivities. A number of atomic transitions with $Q \sim 10^{20}$ 
or higher are being investigated. These include nuclear optical transitions \cite{Campbell2012}, optical transitions in Erbium \cite{Kozlov2013}, transitions in highly charged ions \cite{Derevianko2012}, and the Octupole transition of $^{171}{\rm Yb}^+$ \cite{Huntemann2012}. Such approaches predict a potential improvement of the clock stability to the level of $\sigma_{\rm tomorrow} \sim 10^{-18}/\sqrt{\tau{\rm/sec}}$. Alternatively, trapping more ions would lead to moderate improvements in clock stability and quantum entangling the trapped ions would lead to even further improvement with the ultimate improvement resulting from a network of quantum entangled clocks \cite{Komar2014}. It is therefore not unreasonable to expect measurements in the geoid at sub-millimetre level in the future.  

\subsection{Geopotential Changes}
\begin{figure}[h]
\centering
\includegraphics[width=7cm,clip]{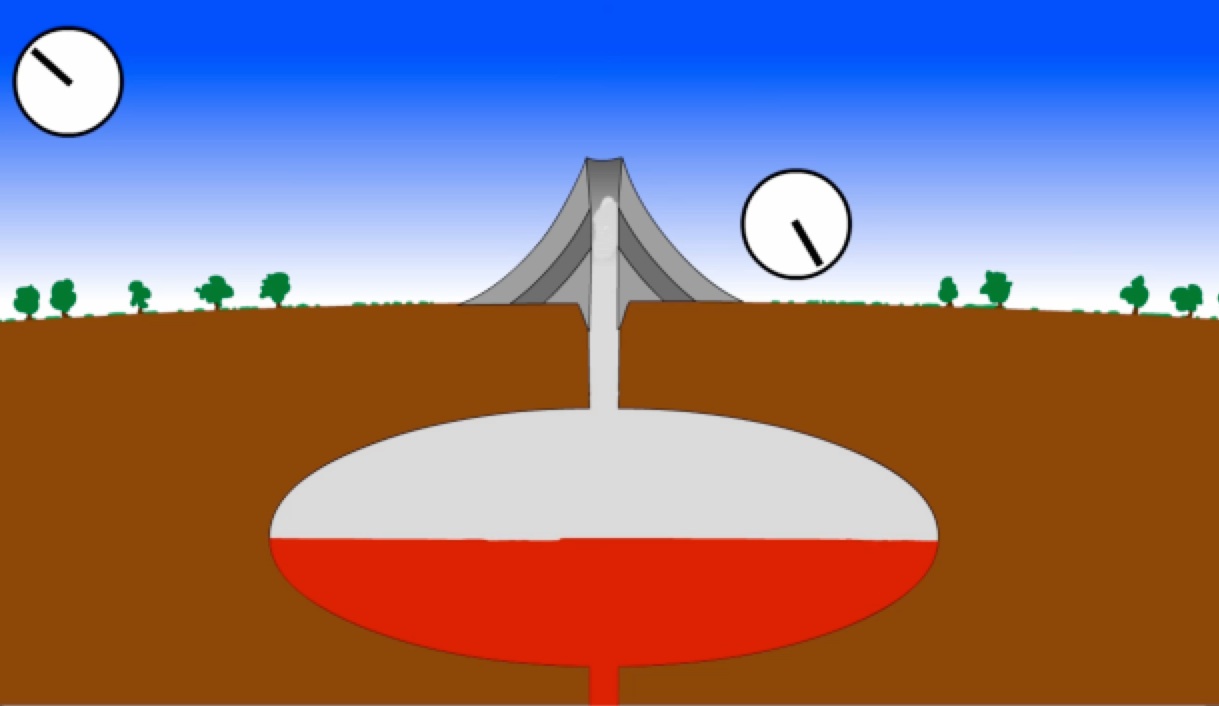}
\caption{A clock placed on a volcano will slow down as the magma chamber fills. A clock further away will maintain its
tick rate.}
\label{fig-1}       
\end{figure}

Atomic clocks measure changes in the geopotential as a function of time at the location of the clock. For a vertical displacement of $h = 1 cm$, 
the geopotential difference can be approximated by
\begin{equation}
\Delta U_{\rm{uplift}} \approx \frac{GM_\oplus}{R_\oplus^2} h \sim 0.1  \left(\frac{h}{1 \, \, \rm cm}\right)\frac{{\rm m}^2}{{\rm sec}^{2}},
\end{equation}
where the clock frequency stability is
\begin{equation}
\frac{\Delta f}{f} \approx \frac{\Delta U}{c^2} \sim 10^{-18}  \left(\frac{h}{1 \, \, \rm cm}\right).
\end{equation}
A spherical magma chamber that fills (See Fig.\ \ref{fig-1}) causes the vertical displacement
\begin{equation}
w({\bf r}) = \frac{d_0}{\left[(r/d)^2 + 1\right]^{3/2}},
\end{equation} 
where $d$ is the distance from the observer to the magma chamber in the vertical direction, $d_0 = (1-\nu) \Delta V_r/(\pi d^2)$, $\Delta V_r = 3 \Delta P V_r/(4 \mu)$, $V_r$ is the volume of the source, $\mu$ is the shear modulus, $\Delta P$ is the pressure change, $\nu$ is $\sim 1/4$,  $r = \sqrt{x^2 + y^2}$, and ${\bf r} = (x, y,0)$ is the location of the observer \cite{gravityref}. 

The dominant term in the potential will then be the vertical uplift with $h = w$
\begin{equation}
\Delta U_{\rm{uplift}} \approx \frac{GM_\oplus}{R_\oplus^2} \frac{d_0}{\left[(r/d)^2 + 1\right]^{3/2}}
\end{equation}
The correction term due to the presence of the density anomaly of mass $\Delta M$ is
\begin{equation}
\Delta U_{\rm{redistr}}  \approx \frac{G \Delta M}{\sqrt{r^2 + d^2}} = \frac{4 \pi G \Delta \rho b^3}{3 \sqrt{r^2 + d^2}},
\end{equation}
where we assume a spherical magma chamber of radius $b$ with density contrast $\Delta \rho$ with the surrounding material.
In addition to this `direct' signal from the presence of the additional magma, there are two other corrections, which have the same
$(r^2 +d^2)^{-1/2}$ dependence and similar magnitude. First, the the inflation of the chamber causes elastic deformation and 
hence density changes of the surrounding rock. The potential signal is
\begin{equation}
\Delta U_{\rm elastic}= \rho_c \frac{(1 - 2 \nu) \Delta V_r}{(r^2+d^2)^{1/2}} ~.
\end{equation}

Second, part of this elastic deformation is a cap of uplifted ground which replaces air,
which is best treated numerically.  The standard approximation of an infinite plane which is made to 
calculate the gravity change yields an infinite answer in the case of the potential.

The term corresponding to mass redistribution is typically 2-3 orders of magnitude lower than that due to vertical uplift for the same
magma intrusion. Thus, unless there is significant magma redistribution that may occur in the case of supervolcanoes, the `direct' signal 
will only be observed if clocks improve to $\sigma_{\rm tomorrow} \sim 10^{-18}/\sqrt{\tau{\rm/sec}}$.
\subsection{Gravity Changes}
The change in the gravitational acceleration caused by the same vertical displacement $h$ is
\begin{equation}
\Delta g_{z \; \rm{uplift}} \approx \frac{2 GM_\oplus}{R_\oplus^3} h \approx 3.1 \left(\frac{h}{1 \, \rm{cm}}\right) \rm{\mu gal},
\end{equation}
where the $z$ subscript is used to emphasise that we have chosen the direction pointing towards the centre of the Earth
of the gravitational acceleration vector. A change in the gravitational acceleration of $\mu {\rm gal}$ is easily measurable
both by relative gravimeters and by quantum gravimeters, which are just being produced commercially. The latter is 
based on atom interferometery that relies on fundamental physics similar to that of atomic clocks, which unlike the 
string of the relative gravimeters should not depreciate in time.

We can set $h = w$ to estimate the gravity change when a magma chamber fills to obtain
\begin{equation}
\Delta g_{z \; \rm{uplift}} \approx  \frac{2 GM_\oplus}{R_\oplus^3} \frac{d_0}{\left[(r/d)^2 + 1\right]^{3/2}}.
\end{equation}
The intruding magma of mass $\Delta M$ has a different density than the rock it displaces, and the direct
signal is proportional to the density contrast and the volume change in the chamber
\begin{equation}
 \Delta g_{z\; \rm{redistr}} \approx  \frac{G \Delta M d}{ (r^2 +d^2)^{3/2}} = \frac{4 \pi G d \Delta \rho b^3}{3 \left(r^2 +d^2\right)^{3/2}}.
\end{equation}
Like before this is the `direct' signal due to the presence of the magma, and the other effects have similar 
dependence on $r$ and $d$.

We have already treated the change in gravity due to the uplift of the gravimeter. An additional effect is the gravity of the uplifted ground 
underneath the gravimeter, which can be modelled as an infinite thin plate displacing air. The result is \cite{gravityref}
\begin{equation}
\Delta g_{z\; \rm{surface}} \approx 2 G \rho_{\rm{c}} \frac{(1- \nu) \Delta V_r d}{\left(r^2 + d^2\right)^{3/2}}
\end{equation}
where $\rho_c$ is the density of the crustal rock. There is also signal due to the density change in the ground caused by the stress of the
intruding magma chamber. This creates a gravity signal of
\begin{equation}
\Delta g_{z\; \rm{density}} \approx - G \rho_c \frac{(1- 2 \nu) \Delta V_r d}{\left(r^2 + d^2\right)^{3/2}} ~.
\end{equation}

We note that in the case of gravimetric measurements there is no way to distinguish between vertical uplift and mass redistribution
since the dependence on the distance to the source is the same. Ideally, one would measure the uplift with a local clock and then
remove that estimate from the gravimetric measurement. Similar analysis has been performed by combining differential GPS with
gravimetric data to determine if any mass redistribution has occurred.
\subsection{The Solid Earth Tide}
A global network of atomic clocks and quantum gravimeters could monitor variations in the amplitude of the solid Earth tide. Such monitoring would be particularly 
valuable in geologically active regions. In this section we discuss the complimentary sensitivity of atomic clocks and gravimeters to the solid Earth tide. 
 Tidal motion occurs because the Earth moves in the gravitational field of the Sun and the Moon. Since the distance to the Sun and the Moon is large relative
  to the size of the Earth, the tidal potential can be written as an expansion about the centre of the Earth \cite{tidalref}
\begin{eqnarray}
V_{\rm tidal} = \frac{GM}{R} \sum_{n=2}^{\infty} \left(\frac{r}{R}\right)^n P_n(\cos\alpha) = \sum_{n=2}^{\infty} V_n(r)
\label{Vtidal}
\end{eqnarray}
where $R$ is the distance from the Sun or the Moon to the centre of the Earth, $r$ is the distance from the centre of the Earth to the observation point, and $\alpha$ is the angle between the lines from the centre of the Earth to the object and the observation point, and the $P_n$ are Legendre polynomials. The sum starts at $n=2$ because the dipole term ($n=1$), which represents a constant force at all points on the Earth, cancels due to the overall acceleration towards either the Sun and the Moon.  Since the gravitational fields are weak, the effects of the Sun and the Moon simply add. 

The Earth responds to this external potential by deforming, which causes an uplift of the ground and a change in the Earth's gravitational potential. As discussed above, atomic clocks are most sensitive to ground uplift. The tidal uplift of the ground $h$ is given by
\begin{equation}
h(r) = \sum_{n=2}^{\infty} \frac{h_n V_n(r)}{g} \approx \frac{h_2 V_2}{g}, 
\end{equation}
where the potential difference measured by atomic clocks is $\Delta V \approx g h$, $h_n$ is a Love number, $g$ is the gravitational acceleration, and $V_n$ is the $n$-th  term of the tidal potential in Eq.\ (\ref{Vtidal}). The dominant term is the
 quadrupole term ($n=2$).  The $n=2$ tide has a semidiurnal period. It can reach an amplitude of up to about $50$~cm at the Equator, which 
corresponds to a $\Delta f/f \sim 5.4 \times 10^{-17} (h/50\,{\rm cm})$ and is measurable by the best atomic clocks over an integration period of about 5 minutes. 
The terms decrease at a rate of $1/60$ at each order for the Moon, and by $4\times 10^{-5}$ with each order for the Sun.

The Love numbers are predicted by various models to 4 or 5 decimal places, but the tidal uplift is not monitored to high precision as a function
of time. Such measurements are challenging with differential GPS because the GPS receivers have to be within a short distance of the order
of $10$ km from a base stations, and the tidal effects cancel.  Atomic clock comparisons on ground could eventually occur with baselines of $10,000$ km, which are comparable to the wavelength of the dominant mode of the solid Earth tide (the ideal comparison at half of its wavelength is with a baseline of a quarter of the circumference of the Earth). They can be used to measure the tidal response of the Earth on the local level and should depend only on factors underground. Very precise measurements of tidal uplift may resolve changes in the deformability of the crust, a possible precursor to earthquakes.


\section{Further Discussion and Conclusions}
It is known that the magnitude of earthquakes and volcanic eruptions is notoriously difficult to predict.
Therefore, using general relativity and atomic clocks to improve such predictions may at first seem
like a bad idea. However, atomic clocks are becoming amazingly precise. They are expected to be delayed
by a fraction of a second in the lifetime of the universe. We used them to measure length and time, and it is
only natural to extend this use to infer mass changes.

Man made activities such as hydraulic fracturing also could benefit from careful monitoring of the solid
Earth tide. Some such activities occur in volcanic areas such as within relative proximity of the Yellowstone 
supervolcano. Calibrating differential GPS data, and performing local clock measurements that are affected 
only by changes underground and not by uncontrollable effects such as variations in the atmosphere could
be important in distinguishing between earthquakes that coincide with mass redistribution and earthquakes
that do not. Additionally, the crust may react differently to tidal deformations in different areas. 
Local monitoring of the solid earth tide in areas of high seismicity can provide the necessary data 
to investigate its potential correlation to earthquakes.






%
%
%

\end{document}